\documentclass[12pt,epsf]{article}
\usepackage{amsmath,amssymb}
\usepackage{graphicx}
\usepackage{cite,fancyhdr}

\usepackage[affil-it]{authblk}
\usepackage[left=2.5cm,right=2.5cm,top=2.5cm,bottom=2.5cm,a4paper]{geometry}

\allowdisplaybreaks

\begin{document}

\setcounter{footnote}{0}
\setcounter{figure}{0}
\setcounter{table}{0}

\title{\bf \Large 
Light Higgsino and Gluino \\ in $R$-invariant Direct Gauge Mediation
}
\author{{\normalsize Ryo Nagai}}
\author{{\normalsize Norimi Yokozaki}}

\affil{\small 
Department of Physics, Tohoku University, \authorcr {\it Sendai, Miyagi 980-8578, Japan}}

\date{}

\maketitle

\thispagestyle{fancy}
\rhead{TU-1051}
\cfoot{\thepage}
\renewcommand{\headrulewidth}{0pt}

\begin{abstract}
\noindent
We provide a simple solution to the $\mu$-$B_\mu$ problem in the ``$R$-invariant direct gauge mediation model".
With the solution, the Higgsino and gluino are predicted to be light as $\mathcal{O}(100)$\,GeV and $\mathcal{O}(1)$\,TeV, respectively. Those gluino and Higgsino can be accessible at the LHC and future collider experiments. Moreover, dangerous dimension five operators inducing rapid proton decays are naturally  suppressed by the $R$-symmetry.

\end{abstract}

\clearpage

\section{Introduction}
Models of gauge mediated supersymmetry breaking (GMSB)~\cite{Dine:1993yw,Dine:1994vc,Dine:1995ag}\,\footnote{
See also Refs.\cite{Dine:1981za,Dimopoulos:1981au,Dine:1981gu,Nappi:1982hm,AlvarezGaume:1981wy} for early attempts.
}
 are very attractive, 
since dangerous flavor violating processes are naturally suppressed: 
soft supersymmetry (SUSY) breaking masses of sleptons and squarks are generated via gauge interactions, and hence, they are flavor-blind.

Among GMSB models, ``$R$-invariant direct gauge mediation model" constructed in Refs.~\cite{Izawa:1997gs,Nomura:1997uu} (see also \cite{Ibe:2012dd, Chiang:2015iva} for recent discussions) is highly successful, since the SUSY breaking minimum is stable. The model has an (spontaneously broken) $R$-symmetry, 
which may suppress dangerous proton decay operators. This model is also interesting from the view point of phenomenology. The gaugino masses are suppressed compared to sfermion masses even though the $R$-symmetry is spontaneously broken. 
These relatively light gauginos can be seen at future large hadron collider (LHC) experiments. On the other hand, squark masses $m_{\tilde q}$ including stop masses are $\mathcal{O}(10)$\,TeV and the observed Higgs boson mass of 125\,GeV~\cite{higgs_mass_atlas,higgs_mass_cms} is easily explained with large radiative corrections from heavy stops~\cite{Okada:1990vk, Okada:1990gg, Ellis:1990nz, Ellis:1991zd, Haber:1990aw}. 

The important remaining issue in this model is the $\mu$\,-\,$B_\mu$ problem~\cite{Dvali:1996cu,Dine:1997qj,Langacker:1999hs,Hall:2002up,Roy:2007nz,Murayama:2007ge,Giudice:2007ca,Liu:2008pa,Csaki:2008sr,DeSimone:2011va}: if $\mu$ and $B_\mu$ terms are generated dynamically, it usually predicts $\mu^2 \ll B_\mu \sim m_{H_{u,d}}^2$, where $m_{H_u}$ ($m_{H_d}$) is a soft SUSY breaking mass for the up-type (down-type) Higgs. 
With the hierarchy of $\mu^2$ and $B_\mu$, 
it has been considered to be difficult to realize the correct electroweak symmetry breaking (EWSB) for $m_{\tilde q}=\mathcal{O}(0.1\,\mathchar`-\,1)$\,TeV. The situation changes for $m_{\tilde q}\gtrsim10$\,TeV, since the hierarchy itself may not be a problem anymore.

The bare $\mu$-term needs to be prohibited. If the bare $\mu$-term is allowed by the $R$-symmetry, 
the dimension five proton decay operators are also allowed by the symmetry under the assumption that the grand unified theory (GUT) exists.\footnote{
If the $R$-charge of $H_u H_d$ is $2$, 
the $R$-charge of the dangerous operator ${\bf 10}\,{\bf 10}\,{\bf 10}\,\bar{\bf 5}$ is also $2$ provided that Yukawa interactions are allowed by the $R$-symmetry.
}
These dimension five operators cause unacceptably rapid proton decays unless the soft SUSY breaking mass scale is extremely high as $\sim 10^{10}$\,GeV~\cite{Dine:2013nga}. The $\mu$\,-\,$B_\mu$ problem might be related to the rapid proton decay problem.

In the minimal GMSB model, 
it has been shown that the $\mu$\,-\,$B_\mu$ problem is solved in a simple and naive way with a slight modification of the GUT relation among messenger masses for $\mu\sim100$\,GeV and $\sqrt{|B_\mu|} \sim m_{\tilde q}\sim10$\,TeV~\cite{Asano:2016oik}.\footnote{
The $\mu$\,-\,$B_\mu$ problem is also solved in a simple way with mini-split SUSY spectra where the stop mass is larger than $\mathcal{O}(100)$\,TeV~\cite{Cohen:2015lyp}.
} 
In this letter, we point out that the $\mu$\,-\,$B_\mu$ problem is also solved in the $R$-invariant direct gauge mediation model in this way.
With the solution, the Higgsino as well as the gluino is predicted to be light, which has a large impact on LHC and International linear collider (ILC) SUSY searches. We also point out that the violation of the GUT relation is not needed for the solution in this model.

\section{$R$-invariant direct gauge mediation and $\mu$/$B_\mu$ term}

\subsection{The model}
First, let us briefly review the $R$-invariant direct gauge mediation model. The model has a spontaneously broken $R$-symmetry, which suppresses gaugino masses compared to sfermion masses. The superpotential of the messenger sector is 
\begin{eqnarray}
W \supset -\mu_Z^2 Z + M_1 \Psi {\bar \Psi}' + M_2 \Psi' {\bar \Psi} + c_1 Z \Psi \bar \Psi, \label{eq:sp1}
\end{eqnarray}
where $\Psi$ and $\Psi'$ ($\bar \Psi$ and ${\bar \Psi}'$) are the messenger fields transformed as $\bf 5$ ($\bf \bar 5$) in $SU(5)$ GUT gauge group. The above superpotential is invariant under $U(1)_R$ symmetry with the $R$-charges of $Q(Z)=Q(\Psi')=Q({\bar \Psi}')=2$ and $Q(\Psi)=Q(\bar \Psi)=0$.
We assume $Z$ has vacuum expectation values, which breaks $R$-symmetry and SUSY as
\begin{eqnarray}
\left<Z\right> = \phi_Z + \left<F_Z\right> \theta^2,
\end{eqnarray}
where $\left<F_Z\right> = \mu_Z^2$. The $R$-symmetry is spontaneously broken by $\left<\phi_Z\right> \neq 0$. 
Such spontaneous breaking of the $R$-symmetry can be achieved in O'Raifeartaigh like models at tree-level~\cite{Carpenter:2008wi, Sun:2008va,Komargodski:2009jf} or one-loop level~\cite{Shih:2007av,Evans:2011pz} if there exists a field with $R$-charge other than 0 or 2. Also, the spontaneously breaking can occur at the higher loop level~\cite{Giveon:2008ne, Intriligator:2008fe,Amariti:2008uz,Amariti:2012pg}.
In this paper, we do not specify the origin of the spontaneous $R$-symmetry breaking and take $\phi_Z$ as a free parameter.

The messenger superfields, $\Psi$, $\Psi'$, $\bar \Psi$ and ${\bar \Psi}'$, are decomposed as 
\begin{eqnarray}
\Psi = \Psi_D + \Psi_{\bar L}, \ 
\bar \Psi = \Psi_{\bar D} + \Psi_L, \  \nonumber \\
\Psi' = \Psi_D' + \Psi_{\bar L}', \ 
{\bar \Psi}' = \Psi_{\bar D}' + \Psi_L', 
\end{eqnarray}
where $\Psi_{\bar D}^{(')}$ and $\Psi_{L}^{(')}$ are transformed as ($\bar{\bf 3}$,\,{\bf 1},\,1/3) and ({\bf 1},\,{\bf 2},\,$-1/2$) under $SU(3)_c \times SU(2)_L \times U(1)_Y$, respectively. 
Then, the superpotential in Eq.(\ref{eq:sp1}) can be written as 
\begin{eqnarray}
W &\supset& -\mu_Z^2 Z + M_{1L} \Psi_{\bar L} \Psi_{L}' + M_{2L} \Psi_{\bar L}' \Psi_{L} + c_L Z \Psi_L \Psi_{\bar L} \nonumber \\
&& + M_{1D} \Psi_{D} \Psi_{\bar D}' +  M_{2D} \Psi_{D}' \Psi_{\bar D} + c_D Z \Psi_D \Psi_{\bar D},
\end{eqnarray}
where all parameters are taken to be real positive without loss of generality. 
For simplicity, further, we take $M_{1L}=M_{2L} \equiv M_L$ and $M_{1D}=M_{2D} \equiv M_D$ in the following discussions.

Accordingly, the messenger sector are parametrized by the following five parameters:
\begin{eqnarray}
\Lambda_{\rm SUSY}, \, M_{\rm mess}, R, \, r_L, R_L
\end{eqnarray}
where $\Lambda_{\rm SUSY}=c_D \mu_Z^2/M_D$, $M_{\rm mess}=M_D$, 
$R= c_D \phi_Z/M_D$,
$r_L=M_L/M_D$ and $R_L=c_L/c_D$. In the case that $c_L=c_D$ and $M_L=M_D$ are satisfied at the GUT scale, 
$r_L$ and $R_L$ are fixed as $r_L \approx R_L \approx 1/1.4$~\cite{Nomura:1997uu}.

After integrating out the messenger fields, gauginos and sfermions obtain soft SUSY breaking masses. The gaugino masses are estimated as
\begin{eqnarray}
M_1 &\simeq& \frac{g_1^2}{16\pi^2} \left( \frac{2}{5} \frac{\Lambda_{\rm SUSY}^3}{M_{\rm mess}^2} \mathcal{F}_D 
+ \frac{3}{5} \frac{\Lambda_{\rm SUSY}^3}{M_{\rm mess}^2} \frac{R_L^3}{r_L^5} \mathcal{F}_L \right),
\nonumber \\
M_2 &\simeq& \frac{g_2^2}{16\pi^2}  \frac{\Lambda_{\rm SUSY}^3}{M_{\rm mess}^2} \frac{R_L^3}{r_L^5} \mathcal{F}_L, \nonumber \\
M_3 &\simeq& \frac{g_3^2}{16\pi^2}  \frac{\Lambda_{\rm SUSY}^3}{M_{\rm mess}^2} \mathcal{F}_D, \label{eq:gaugino}
\end{eqnarray}
where $\mathcal{F}_D$ and $\mathcal{F}_{L}$ are numerical coefficient of $\mathcal{O}(0.1)$ (see \cite{Nomura:1997uu} for complete formulae). Note that the gaugino masses are suppressed by factors, $(\Lambda_{\rm SUSY}/M_{\rm mess})^2$ and $\mathcal{F}_{L,D}$. On the other hand, the sfermion masses are not suppressed by the factor, and they are approximately given by
\begin{eqnarray}
\tilde m_{i}^2 \simeq \frac{2}{(16\pi^2)^2} \left[ C_3^i \, g_3^4 +  C_2^i \, g_2^4 \frac{R_L^2}{r_L^2} + \frac{3}{5} g_1^4 (Q_Y^i)^2 \left( 
\frac{2}{5} + \frac{3}{5}\frac{R_L^2}{r_L^2}
\right)
\right]  \Lambda_{\rm SUSY}^2, \label{eq:sfermion}
\end{eqnarray}
where $C_3^i (C_2^i)$ is a quadratic Casimir invariant of $SU(3)_c$ ($SU(2)_L$) and $Q_Y^i$ is a hyper charge.
From Eq.(\ref{eq:gaugino}) and Eq.(\ref{eq:sfermion}), we see the hierarchical masses of $M_{1,2,3}^2 \ll \tilde{m}_i^2$.  The complete formula of Eq.(\ref{eq:sfermion}) can be found in, for instance, Refs.~\cite{Marques:2009yu,Sato:2009dk}.

\subsection{Generation of $\mu$/$B_\mu$ terms}

Next, we introduce messenger-Higgs couplings to generate $\mu$ and $B_\mu$-terms.
The relevant part of the superpotential is given by
\begin{eqnarray}
W \supset  c_S Z \mathcal{S} \bar{\mathcal{S}} + k_u H_u \Psi_L \mathcal{S}
+  k_d H_d \Psi_{\bar L} \bar{\mathcal{S}},
\end{eqnarray}
where $\mathcal{S}$ and $\bar{\mathcal{S}}$ are gauge singlet superfields with the $R$-charge assignment, $Q(\mathcal{S})+Q(\bar{\mathcal{S}})=0$. Here, $Q(H_u)+Q(H_d)=4$
and the bare $\mu$ term, $\mu H_u H_d$, is not allowed by $U(1)_R$ symmetry.\footnote{
With a particular choice of $R$-charges, the seesaw mechanism can be incorporated.} 
Also, a dangerous dimension five proton decay operator, ${\bf 10\,10\,10\,\bar 5}$, is prohibited by the symmetry.
So far we have eight free parameters in this model:
\begin{eqnarray}
\Lambda_{\rm SUSY}, \, M_{\rm mess}, R, r_L, R_L,  R_S, k_u, k_d,
\end{eqnarray}
where $\Lambda_{\rm SUSY}$, $M_{\rm mess}$, 
$R$, $r_L$ and $R_L$ are defined in the previous subsection, and $R_S=c_S/c_D$.

Integrating out the messenger fields, $\mathcal{S}$ and $\bar{\mathcal{S}}$, the $\mu$-parameter and soft SUSY breaking mass parameters are generated as 
\begin{eqnarray}
\mu &\approx& -160\, {\rm GeV} \left(\frac{k_u k_d}{0.05}\right) 
\left(\frac{\Lambda_{\rm SUSY}}{2 \cdot 10^6{\rm GeV}}\right) \nonumber \\
B_\mu &\approx& 1.2 \times 10^8\,  {\rm GeV}^2 \left(\frac{k_u k_d}{0.05}\right) 
\left(\frac{\Lambda_{\rm SUSY}}{2 \cdot 10^6{\rm GeV}}\right)^2 \nonumber \\
\delta m_{H_{u,d}}^2 &\approx& 9.8 \times 10^8 \, {\rm GeV}^2 \left(\frac{k_{u,d}}{0.5} \right)^2
\left(\frac{\Lambda_{\rm SUSY}}{2 \cdot 10^6{\rm GeV}}\right)^2 \nonumber \\
A_{u,d} &\approx& 2.8 \times 10^3\,  {\rm GeV} \left(\frac{k_{u,d}}{0.5} \right)^2
\left(\frac{\Lambda_{\rm SUSY}}{2 \cdot 10^6{\rm GeV}}\right) \label{eq:mubmu}
\end{eqnarray}
for $R=r_L=R_L=1$, $R_S=7$ and $\Lambda_{\rm SUSY}/M_{\rm mess}=0.95$. 
The analytic forms of Eq.(\ref{eq:mubmu}) can be found in Appendix A.

The above $\mu$ and $B_\mu$ must satisfy conditions for the EWSB. The conditions are given by
\begin{eqnarray}
\frac{m_Z^2}{2} &\simeq& \left[ -\mu^2 
- \frac{(m_{H_u}^2  + \frac{1}{2 v_u}\frac{\partial \Delta V}{\partial v_u} ) \tan^2\beta}{\tan^2\beta-1} 
 + \, \frac{m_{H_d}^2 + \frac{1}{2 v_d}\frac{\partial \Delta V}{\partial v_d} }{\tan^2\beta-1} \right]_{M_{\rm stop}}, \nonumber \\
\frac{B_\mu \,(\tan^2\beta+1)}{\tan\beta} &\simeq& \left[ m_{H_u}^2 +\frac{1}{2 v_u}\frac{\partial \Delta V}{\partial v_u} + m_{H_d}^2  + \frac{1}{2 v_d}\frac{\partial \Delta V}{\partial v_d} + 2\mu^2 \right]_{M_{\rm stop}}, \label{eq:ewsb}
\end{eqnarray}
where $m_Z$ is the $Z$ boson mass and $\tan\beta$ is a ratio of the VEVs, $v_u/v_d$; $\Delta V$ is one-loop corrections to the Higgs potential. The Higgs soft masses and $\Delta V$ are evaluated at the stop mass scale, $M_{\rm stop}$. 

The $\mu$-parameter is roughly estimated as
\begin{eqnarray}
-\mu^2 &\simeq& m_{H_u}^2 (M_{\rm stop})+ \frac{1}{2 v_u}\frac{\partial \Delta V}{\partial v_u} - \frac{m_{H_d}^2(M_{\rm stop})}{\tan^2\beta} \nonumber \\
&\simeq& (m_{H_u}^2)_{\rm GMSB} + \delta m_{H_u}^2 + (\Delta m_{H_u}^2)_{\rm rad} + \frac{1}{2 v_u}\frac{\partial \Delta V}{\partial v_u} \nonumber \\
&-& \frac{(m_{H_d^2})_{\rm GMSB} + \delta m_{H_d}^2}{\tan^2\beta}
= -(\mathcal{O}(100)\,{\rm GeV})^2,
\end{eqnarray}
where $(m_{H_{u,d}}^2)_{\rm GMSB}$ are contributions from gauge mediation in Eq.(\ref{eq:sfermion}), and 
$(\Delta m_{H_u}^2)_{\rm rad}$ contains radiative corrections from stop and gluino loops and is negative. 
Since $|(m_{H_u}^2)_{\rm GMSB}| \ll |(\Delta m_{H_u}^2)_{\rm rad}|$, $\mu$-parameter determined by the EWSB conditions is larger than $\mathcal{O}(0.1) M_{\rm stop}$ in usual GMSB models. However, in our model, the small $\mu$-parameter is obtained with sizable $\delta m_{H_u}^2$, i.e. Eq.(\ref{eq:mubmu}) and Eq.(\ref{eq:ewsb}) are consistently satisfied.

\section{Results}
In this section, we discuss the mass spectra of SUSY particles and survey the parameter region where the mass of observed Higgs boson and the EWSB are correctly explained. 
\subsection{SM-like Higgs mass}
\begin{figure}[t]
 \begin{center}
   \includegraphics[width=70mm]{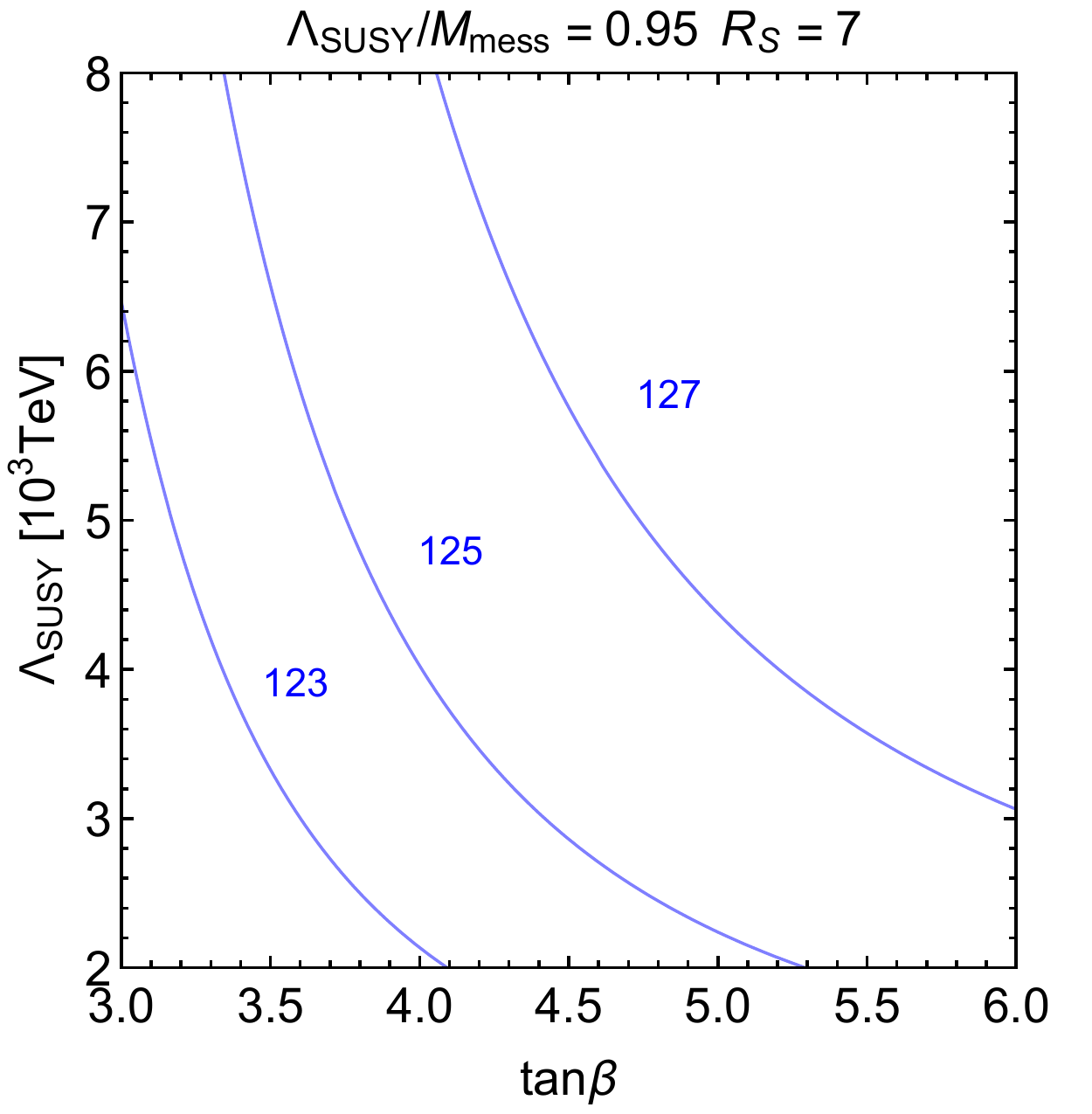}
   ~
   \includegraphics[width=70mm]{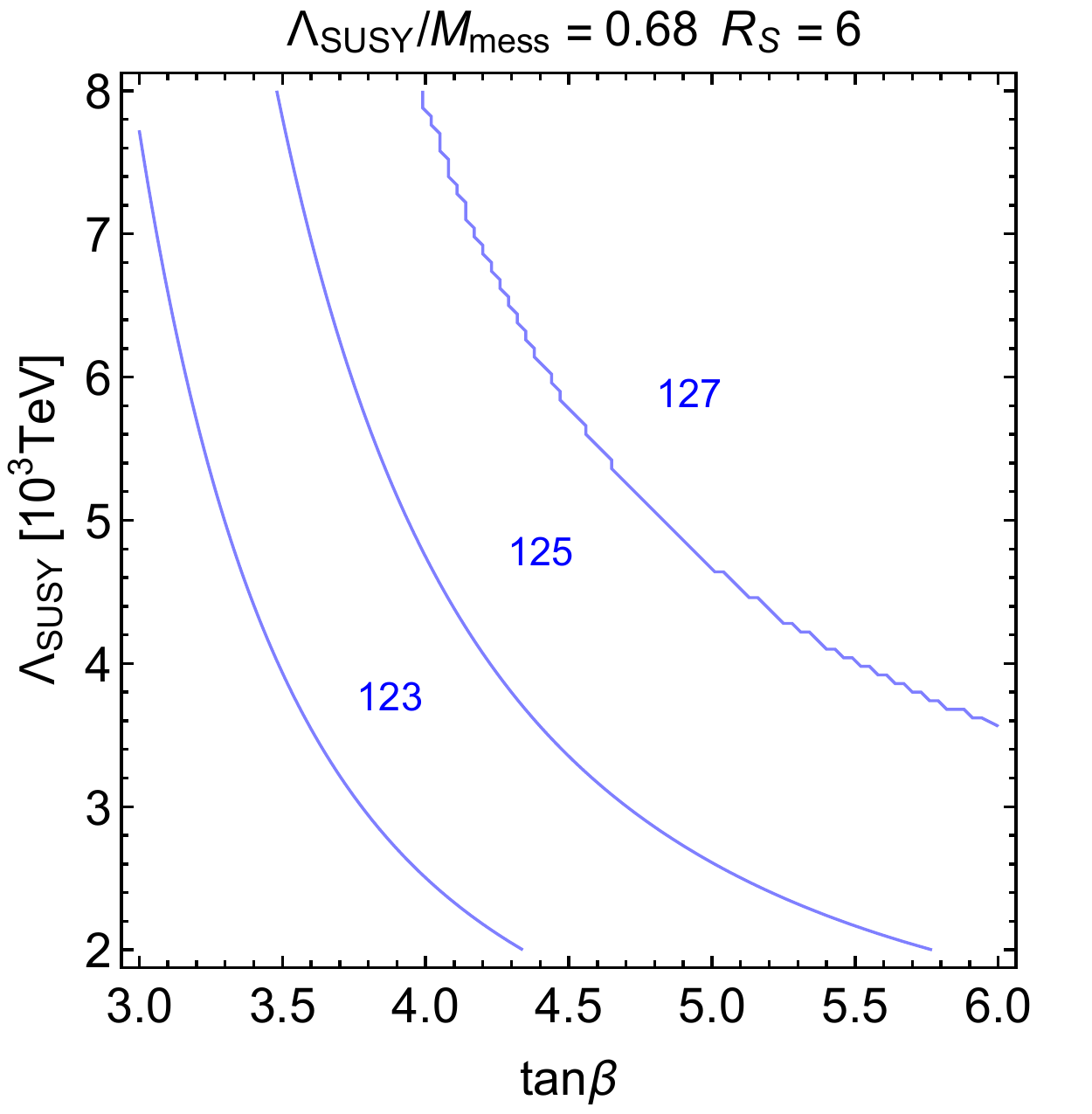}
 \end{center}
  \caption{The SM-like Higgs mass in $(\Lambda_{\rm{SUSY}}, \tan\beta)$ plane. We take $\Lambda_{\rm{SUSY}}/M_{\rm{mess}}=0.95$, $R_L=r_L=1$, $R_S=7$, $k_u=0.1$ and $k_d=0.5$ ($\Lambda_{\rm{SUSY}}/M_{\rm{mess}}=0.68$, $R=1.5$, $R_L=r_L=1/1.4$, $R_S=6$, $k_u=0.02$ and $k_d=0.2$) in the left (right) figure. In both cases,  we take $\alpha_s(M_Z)=0.1185$ and $m_t ({\rm pole})=173.3$GeV.}
 \label{fig:mhiggs}
\end{figure}

First, we estimate the mass of the lightest CP-even neutral Higgs boson, $m_{h^0}$.  Figure~\ref{fig:mhiggs} shows the value of $m_{h^0}$ on $(\Lambda_{\rm{SUSY}}, \tan\beta)$ plane with the other parameters fixed. 
Here we compute mass spectra of SUSY particles using {\tt{softsusy-4.0.1}} \cite{Allanach:2001kg} with appropriate modifications and then $m_{h^0}$ is estimated using {\tt{SUSYHD}} \cite{Vega:2015fna}. 
In the left (right) figure, we take $\Lambda_{\rm{SUSY}}/M_{\rm{mess}}=0.95$, $R_L=r_L=1$, $R_S=7$, $k_u=0.1$ and $k_d=0.5$ ($\Lambda_{\rm{SUSY}}/M_{\rm{mess}}=0.68$, $R=1.5$, $R_L=r_L=1/1.4$, $R_S=6$, $k_u=0.02$ and $k_d=0.2$). In the estimation, $\alpha_s(M_Z)$ and $m_t ({\rm pole})$ are taken to be $\alpha_s(M_Z)=0.1185$ and $m_t({\rm pole})=173.3$\,GeV. Three blue lines in the figure correspond to $m_{h^0}=123, 125$, and $127$\,GeV, respectively. We find that the observed Higgs boson mass of $125$GeV is explained for $\Lambda_{\rm{SUSY}}=\mathcal{O}(10^3)$\,TeV leading to $\mathcal{O}(10)$\,TeV squarks.

\subsection{Gluino mass}
\begin{figure}[t]
 \begin{center}
   \includegraphics[width=70mm]{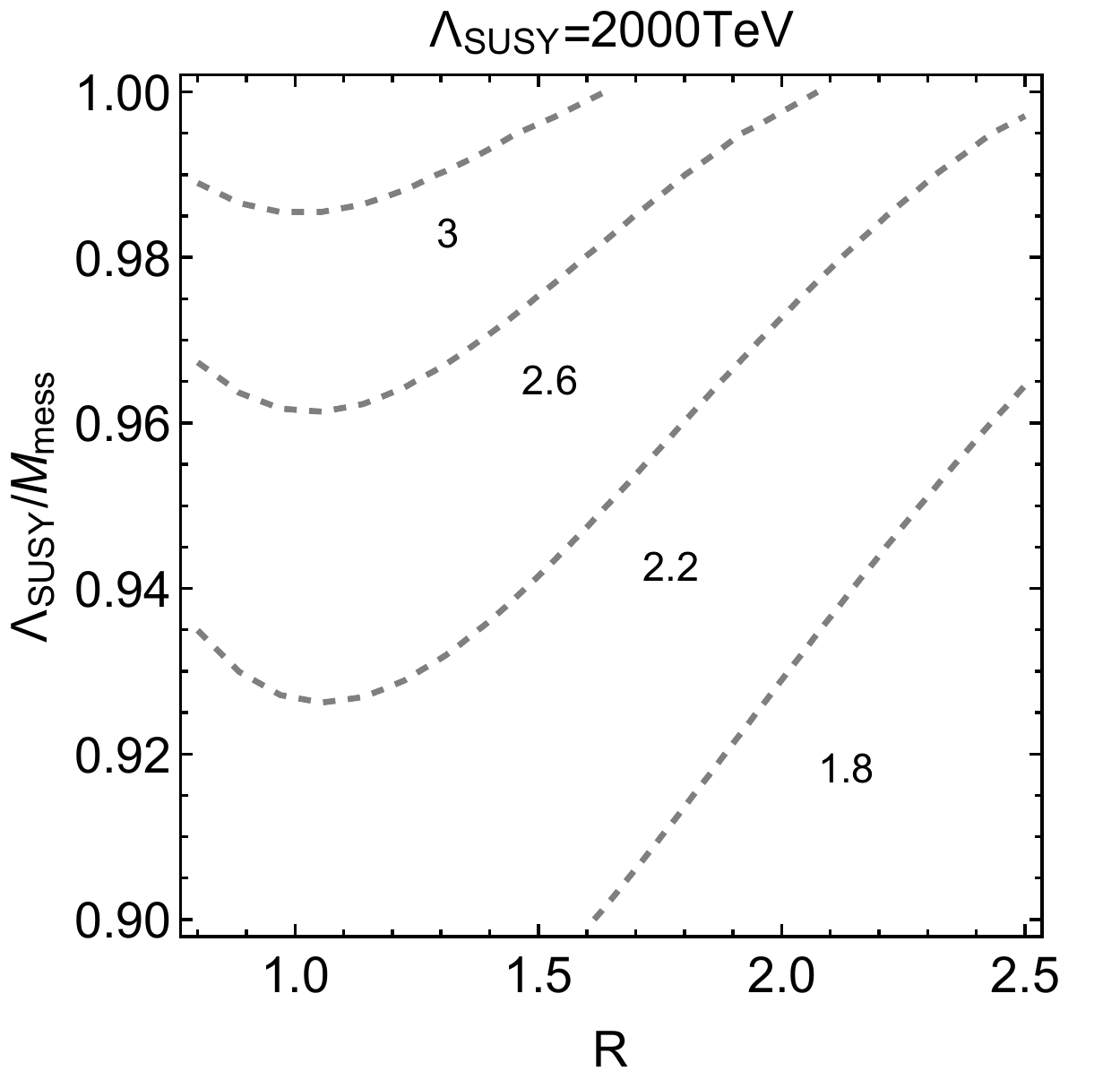}
     ~
   \includegraphics[width=70mm]{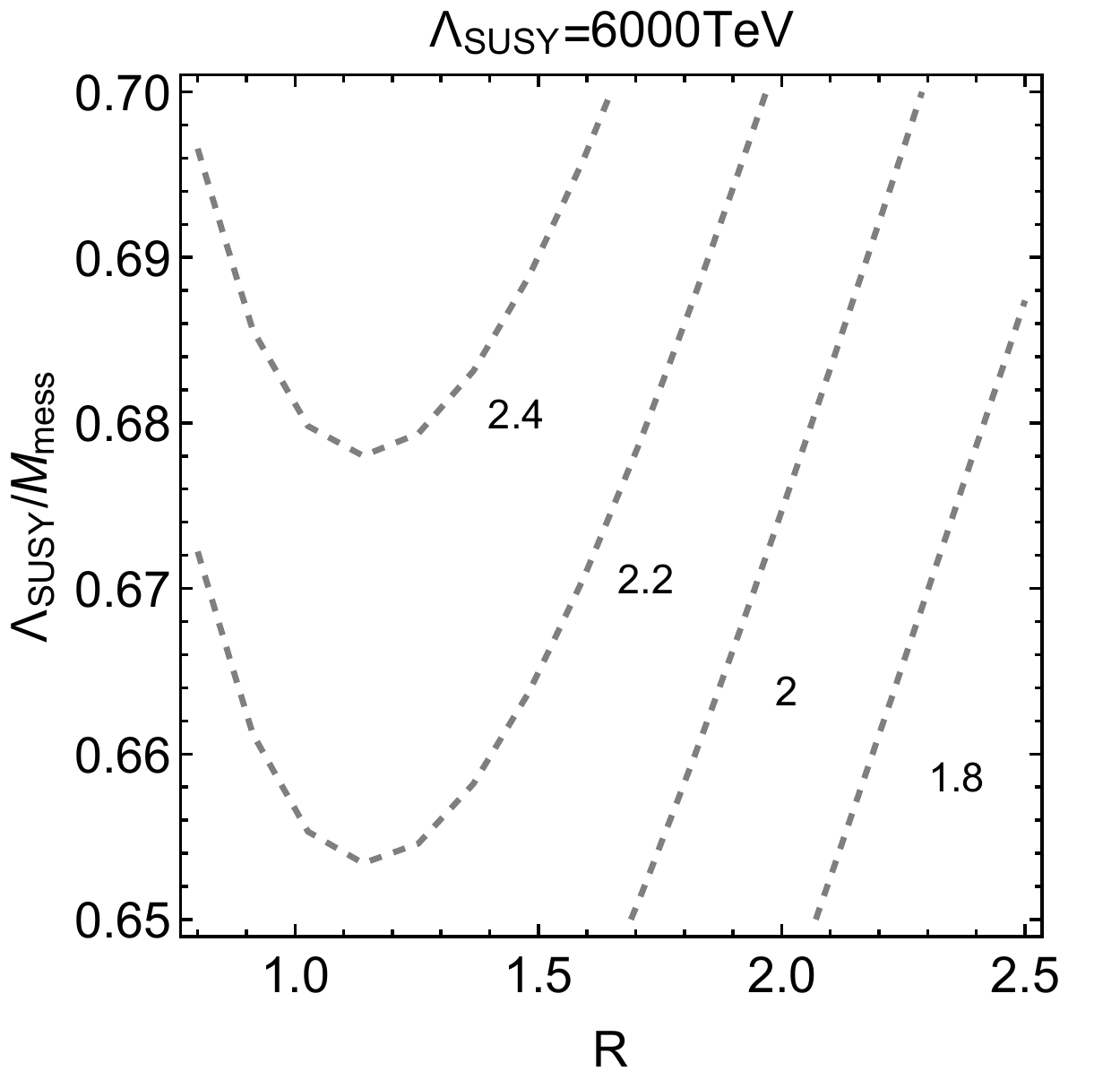}
 \end{center}
  \caption{Gluino mass in $(\Lambda_{\rm{SUSY}}/M_{\rm{mess}}, R)$ plane. We take $\Lambda_{\rm{SUSY}}=2000$TeV, $R_L=r_L=1$, $R_S=10$, $k_u=0.07$ and $k_d=0.28$ ($\Lambda_{\rm{SUSY}}=6000$TeV, $R_L=r_L=1/1.4$, $R_S=6$, $k_u=0.02$ and $k_d=0.2$) in the left (right) figure.}
 \label{fig:mgluino}
\end{figure}

Even for $\Lambda_{\rm{SUSY}} = \mathcal{O}(10^3)$\,TeV (i.e.\,$\mathcal{O}(10)$\,TeV squarks), the gauginos in our model are predicted to be enough light for good targets at the collider experiments. Figure~\ref{fig:mgluino} shows the mass of gluino, $m_{\tilde{g}}$, on $(\Lambda_{\rm{SUSY}}/M_{\rm{mess}}, R)$ plane with fixing $\Lambda_{\rm{SUSY}}$ and the other parameters. For the estimation of $m_{\tilde{g}}$, we use {\tt{softsusy-4.0.1}}. In the left (right) figure, we take $\Lambda_{\rm{SUSY}}=2000$\,TeV, $R_L=r_L=1$, $R_S=10$, $k_u=0.07$ and $k_d=0.28$ ($\Lambda_{\rm{SUSY}}=6000$\,TeV, $R_L=r_L=1/1.4$, $R_S=6$, $k_u=0.02$ and $k_d=0.2$). Gray dashed lines show the contours of $m_{\tilde{g}}$ [TeV]. It is found that $m_{\tilde{g}} = 2$\,-\,3\,TeV in the whole parameter region shown in Fig.~\ref{fig:mgluino}.

\begin{figure}[t]
 \begin{center}
   \includegraphics[width=72mm]{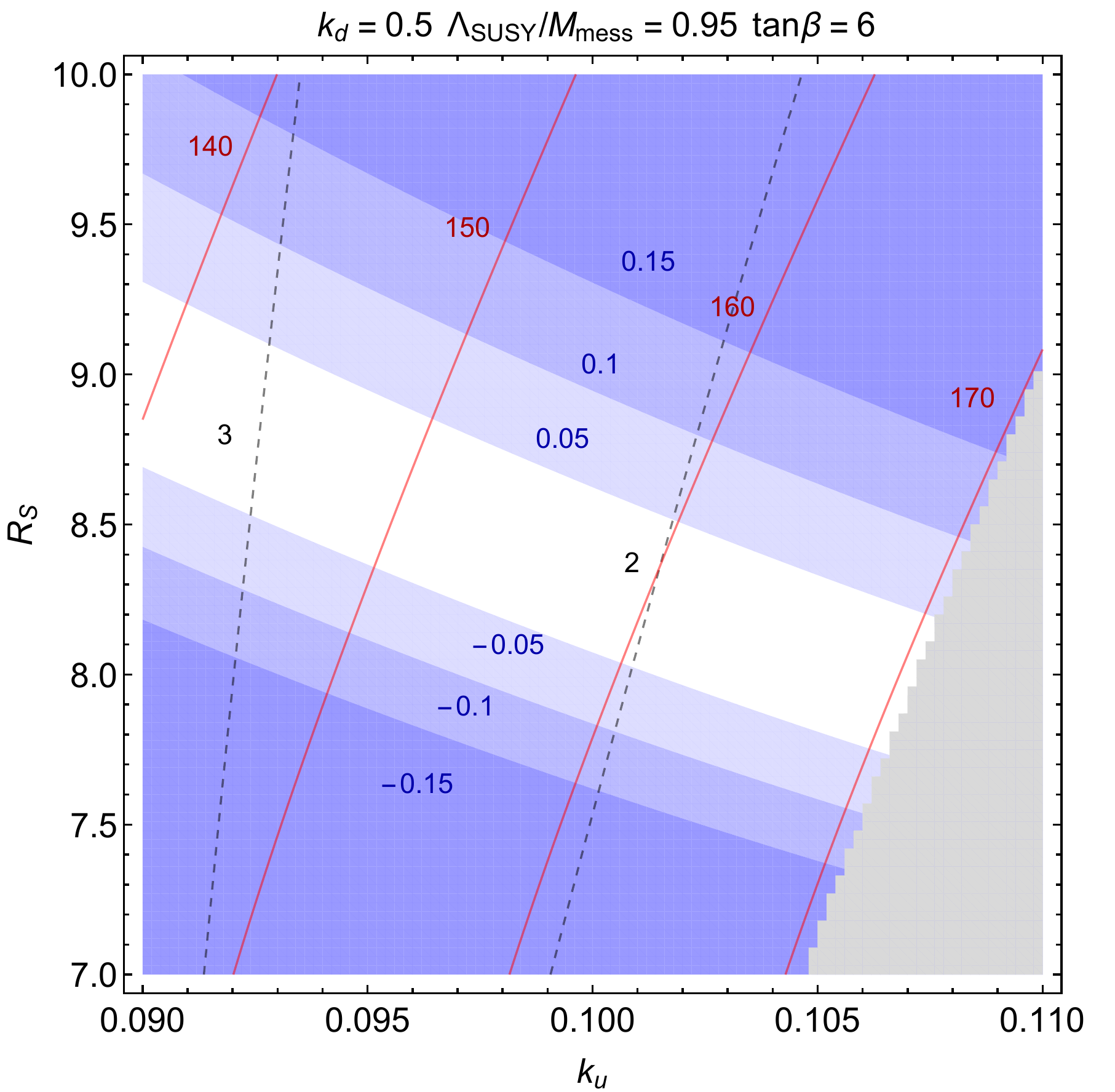}
   ~
   \includegraphics[width=70mm]{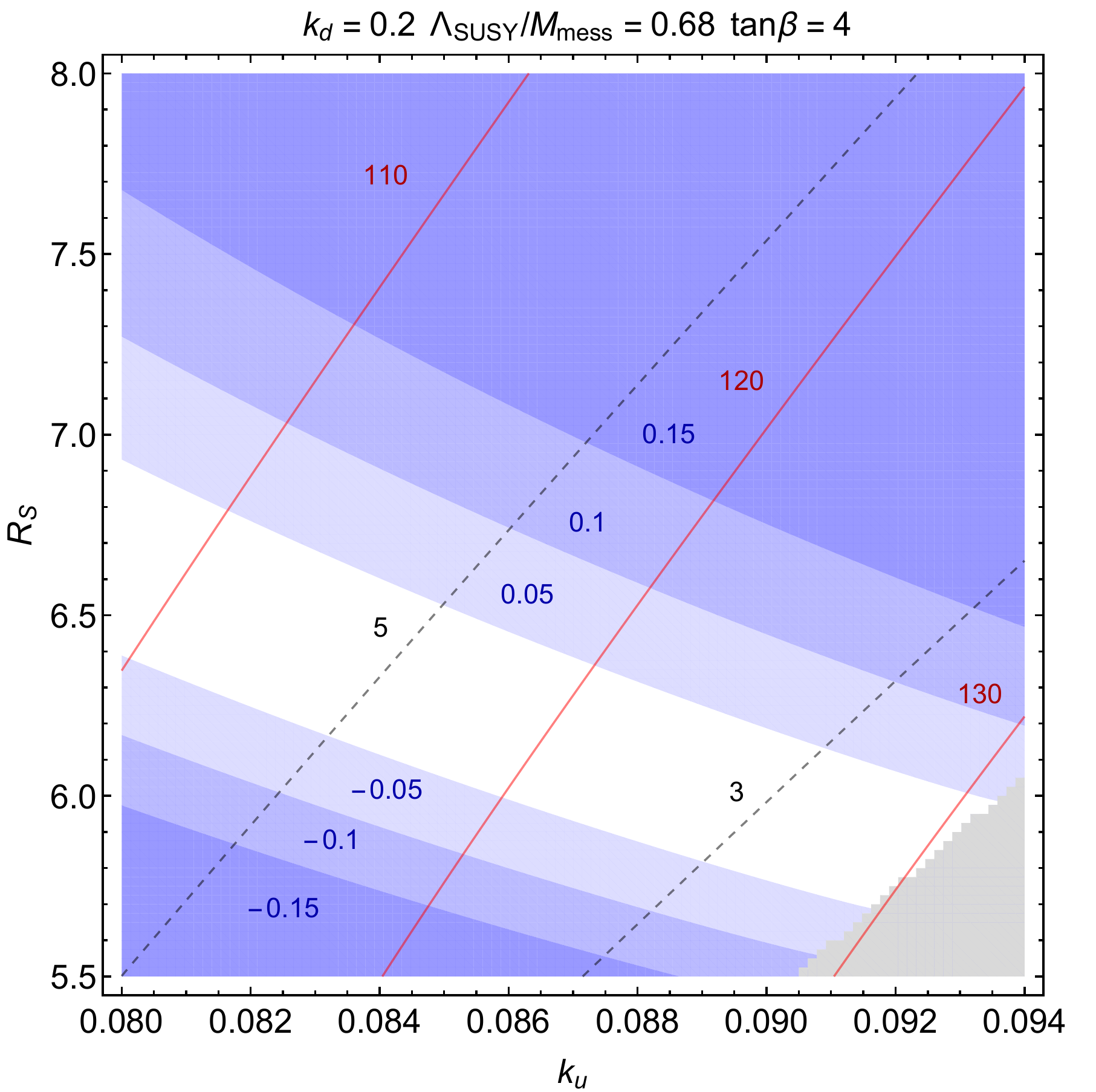}
 \end{center}
  \caption{$\delta B_{\mu}=(|B_{\mu}|-|B^{\rm{EWSB}}_{\mu}|)/|B_{\mu}|$ (Blue region), the contour of $|\mu|$[GeV] (red lines) and $|\mu^{\rm{EWSB}}|$[TeV] (black dashed lines). We take $\Lambda_{\rm{SUSY}}/M_{\rm{mess}}=0.95$, $R=R_L=r_L=1$, and $k_d=0.5$ ($\Lambda_{\rm{SUSY}}/M_{\rm{mess}}=0.68$, $R=1.5$, $R_L=r_L=1/1.4$, and $k_d=0.2$) in the left (right) figure.}
 \label{fig:EWSB}
\end{figure}

\begin{table}[t]
\begin{center}
\caption{Mass spectra in some benchmark points. Mass spectra of SUSY particles and the SM like Higgs boson are computed using {\tt{softsusy-4.0.1}} and {\tt{SUSYHD}}, respectively. At all benchmark points, we take $\alpha_s(M_Z)=0.1185$ and $m_t$(pole)$=173.3$GeV.}
\vspace{5pt}
\begin{tabular}{|c||c|c|c|c|}
\hline 
Parameter & Point (A) & Point (B) & Point (C) & Point (D) \\
\hline
$\Lambda_{\rm{SUSY}}$ & 2000\mbox{TeV} & 4000\mbox{TeV} & 6000\mbox{TeV} & 6000\mbox{TeV} \\
$\Lambda_{\rm{SUSY}}/M_{\rm{mess}}$ & 0.95 & 0.80 & 0.68  & 0.68 \\
$R$ & 1 & 1 & 1.5 & 1.8\\
$R_L$ & 1 & 1 & 1/1.4 & 1/1.4 \\
$r_L$ & 1 & 1 & 1/1.4 & 1/1.4 \\
$R_S$ & 8 & 7 & 6 & 6.8 \\
$k_u$ & $\approx0.11$ & $\approx0.10$ & $\approx0.09$ & $\approx0.16$\\
$k_d$ & 0.5 & 0.3 & 0.2 & 0.5 \\
\hline
Prediction & & & & \\
\hline
$m_{\tilde{g}}$ & 2.45\mbox{TeV} & 2.53\mbox{TeV} & 2.53\mbox{TeV} & 2.16\mbox{TeV} \\
$m_{\tilde{\chi}^0_1}$ & 162\mbox{GeV} & 180\mbox{GeV} & 128\mbox{GeV} & 455\mbox{GeV} \\
$m_{\tilde{\chi}^0_2}$ & 186\mbox{GeV} & 200\mbox{GeV} & 138\mbox{GeV} & 465\mbox{GeV} \\
$m_{\tilde{\chi}^0_3}$ & 424\mbox{GeV} & 539\mbox{GeV} & 861\mbox{GeV}& 989\mbox{GeV} \\
$m_{\tilde{\chi}^0_4}$ & 802\mbox{GeV} & 888\mbox{GeV} & 2.20\mbox{TeV} & 2.17\mbox{TeV} \\
$m_{\tilde{\chi}^\pm_1}$ & 174\mbox{GeV} & 190\mbox{GeV} & 132\mbox{GeV} & 460\mbox{GeV} \\
$m_{\tilde{\chi}^\pm_2}$ & 790\mbox{GeV} & 877\mbox{GeV} & 2.20\mbox{TeV}& 2.16\mbox{TeV} \\
$(m_{\tilde{t}_1}$,$m_{\tilde{t}_2})$ & (12.9, 14.7)\mbox{TeV}  & (25.5, 28.9)\mbox{TeV} & (32.8, 36.9)\mbox{TeV} & (29.6, 33.6)\mbox{TeV} \\
$(m_{\tilde{e}_L}$,$m_{\tilde{e}_R})$ & (5.39, 3.92)\mbox{TeV}  & (11.4, 7.04)\mbox{TeV} & (14.7, 8.37)\mbox{TeV} & (13.1, 8.47)\mbox{TeV} \\
$m_{h^0}$ & 125.7\mbox{GeV}  & 125.6\mbox{GeV}  & 125.7\mbox{GeV} & 125.5\mbox{GeV}  \\
$m_{A^0}$ & 30.1\mbox{TeV} & 40.9\mbox{TeV} & 28.4\mbox{TeV} & 48.5\mbox{TeV}  \\
\hline
$\mu$ & -171\mbox{GeV} & -187\mbox{GeV} & -130\mbox{GeV} & -459\mbox{GeV} \\
$\tan\beta$ & 6 & 4 & 4 & 4 \\
\hline
\end{tabular} \label{tab:points}
\end{center}
\end{table}

\subsection{Electroweak symmetry breaking}
We next check whether the EWSB conditions are correctly satisfied. For this purpose, we solve the EWSB conditions (Eq.~(\ref{eq:ewsb})) using {\tt{softsusy-4.0.1}} and compare the solutions with  $\mu$ and $B_\mu$ in Eq.(\ref{eq:mubmu}). Figure~\ref{fig:EWSB} shows the difference between our predictions ($\mu$ and $B_\mu$) and the solution to the EWSB conditions ($\mu^{\rm{EWSB}}$ and $B_\mu^{\rm{EWSB}}$). In the left (right) figure, we take $\Lambda_{\rm{SUSY}}/M_{\rm{mess}}=0.95$, $R=R_L=r_L=1$, and $k_d=0.5$ ($\Lambda_{\rm{SUSY}}/M_{\rm{mess}}=0.68$, $R=1.5$, $R_L=r_L=1/1.4$, and $k_d=0.2$). In gray region, $\mu_{\rm EWSB}^2 < 0$. The blue region corresponds to the parameter region where $\delta B_{\mu}=(|B_{\mu}|-|B^{\rm{EWSB}}_{\mu}|)/|B_{\mu}|$ is larger than $0.05$, $0.10$, $0.15$ or smaller than $-0.05$, $-0.10$, $-0.15$. In white region, the prediction of $B_{\mu}$ is consistent with the EWSB conditions within $5$\% level. Red and black dashed lines show the contours of $|\mu|$[GeV] and $|\mu^{\rm{EWSB}}|$[TeV], respectively. It should be noted that the difference between $|\mu|$ and $|\mu^{\rm{EWSB}}|$ is very sensitive to $k_u$. 
In other words, we need a fine-tuning of $k_u$ to find the parameter region where the prediction of ${\mu}$ is consistent with the EWSB conditions.

\subsection{Mass spectra in some benchmark points}
Finally, we show the typical mass spectra in our model. Here, we pick up four benchmark points shown in Table \ref{tab:points}: (A) $\Lambda_{\rm{SUSY}}=2000$\,TeV,~$\Lambda_{\rm{SUSY}}/M_{\rm{mess}}=0.95$,~$R=R_L=r_L=1$,~$R_S=8$,~$k_u\approx 0.11$, and $k_d=0.5$. (B) $\Lambda_{\rm{SUSY}}=4000$\,TeV,~$\Lambda_{\rm{SUSY}}/M_{\rm{mess}}=0.80$,~$R=R_L=r_L=1$,~$R_S=7$,~$k_u\approx 0.10$, and $k_d=0.3$. (C) $\Lambda_{\rm{SUSY}}=6000$\,TeV,~$\Lambda_{\rm{SUSY}}/M_{\rm{mess}}=0.68$,~$R=1.5$,~$R_L=r_L=1/1.4$,~$R_S=6$,~$k_u\approx 0.09$, and $k_d=0.2$. (D) $\Lambda_{\rm{SUSY}}=6000$\,TeV,~$\Lambda_{\rm{SUSY}}/M_{\rm{mess}}=0.68$,~$R=1.8$,~$R_L=r_L=1/1.4$,~$R_S=6.8$,~$k_u\approx 0.16$, and $k_d=0.5$. 
At all benchmark points, we check that the EWSB conditions are correctly satisfied. 
Note that benchmark points (C) and (D) respect GUT relation ($R_L=r_L=1/1.4$). Thus, our model solves $\mu$-$B_\mu$ problem without violating GUT relation. 
It should also be noted that, although the sfermions and extra Higgs bosons are rather heavy, gaugino and Higgsino are always light. The typical masses of Higgsino and gaugino are $\mathcal{O}(100)$GeV and $\mathcal{O}(1)$TeV and they can be good targets for the forthcoming collider experiments.

In our model, the lightest SUSY particle (LSP) is always gravitino. Typical gravitino mass is estimated as
\begin{eqnarray}
m_{3/2}
=\frac{\mu^2_Z}{\sqrt{3} M_{\rm{pl}}} \approx 10\,{\rm keV}\left(\frac{0.1}{c_D}\right) 
\left(\frac{\Lambda_{\rm SUSY}}{2000\,{\rm TeV}} \right)^2
\left(\frac{0.95}{\Lambda_{\rm SUSY}/M_{\rm mess}}\right),
\end{eqnarray}
where ${M_{\rm{pl}}}\simeq 2.4\times 10^{18}$GeV denotes the reduced Planck mass.{\footnote{
Provided that the $R$-symmetry is explicitly broken by a constant term in the superpotential, the mass of the $R$-axion is given by
\begin{eqnarray}
m_{a}\simeq 
8.4\,\mbox{GeV}\left(\frac{m_{3/2}}{10\,\mbox{keV}}\right)\left(\frac{9000\,\mbox{TeV}}{\phi_Z}\right)^{\frac{1}{2}}.
\end{eqnarray}
}} 
With the above gravitino mass, the next to the lightest SUSY particle behaves as a stable particle in collider time scale.

\section{Conclusion and discussion}

We have provided a simple solution to the $\mu$-$B_\mu$ problem in $R$-invariant direct gauge mediation. 
In contrast to the case of minimal gauge mediation shown in Ref.~\cite{Asano:2016oik}, 
the solution works even when the GUT relations among the parameters in the messenger sector are satisfied. 
 
The Higgsino is predicted to be light as $\sim100$\,-\,$500$\,GeV with the solution. 
Since the gravitino is expected to be heavier than 10\,-\,100\,keV, the lightest neutralino, which is Higgsino-like, is stable 
inside a detector. This light Higgsino is a good target at the LHC~\cite{Han:2013usa,Schwaller:2013baa,Baer:2014cua,Han:2014kaa,Han:2015lma,Barducci:2015ffa} and ILC~\cite{Berggren:2013vfa}. 
The gluino is also likely to be light as 2\,-\,3\,TeV, which can be tested at the future LHC experiment~\cite{14tev_atlas}. 
Moreover, the dangerous dimension five operators inducing rapid proton decays are naturally suppressed by the $R$-symmetry.

\section*{Acknowledgments}
We thank Yuji Omura for useful discussions.
This work is supported by JSPS KAKENHI Grant Numbers JP15H05889 (N.Y.), JP15K21733 (N.Y.), JP17H05396 (N.Y.), JP17H02875 (N.Y.), and 16H06490 (R.N.).


\appendix

\section{Analytic formulae for $\mu/B_\mu$-term, $A$-terms and $m_{H_{u,d}}^2$} \label{app:formulae}
In this appendix, we give analytic formulae for $\mu/B_\mu$-term, $A$-terms and $m_{H_{u,d}}^2$ at one-loop level. The definition for these parameters is the same with that in Ref.~\cite{Asano:2016oik}. 

To begin with, we summarize the mass eigenstates of the messenger fermions and sfermions. After the spontaneous SUSY and $U(1)_R$ symmetry breaking, the mass matrices for messenger lepton and slepton, ${\bf{m}}_L$ and $\widetilde{{\bf{m}}}^2_L$, are given by
\begin{align}
{\bf{m}}_L
=
\left(
\begin{array}{ccc}
c_L\phi_Z & M_{2L}\\
M_{1L} & 0\\
\end{array}
\right),
~~~
\widetilde{{\bf{m}}}^2_L
=
\left(
\begin{array}{cc}
{\bf{m}}^T_L{\bf{m}}_L & -{\bf{c}}_L\mu^2_Z\\
-{\bf{c}}_L\mu^2_Z & {\bf{m}}_L{\bf{m}}^T_L
\end{array}
\right),
~~~
{\bf{c}}_L
=
\left(
\begin{array}{cc}
c_L & 0\\
0 & 0
\end{array}
\right).
\end{align}
These mass matrices are diagonalized by orthogonal matrices $U$, $V$ and ${\widetilde{V}}$ as
\begin{align}
&U^T{\bf{m}}_L V
=
\mbox{diag}(m_{L_1},m_{L_2}),\\
&{\widetilde{V}}^T{\widetilde{{\bf{m}}}^2_L}{\widetilde{V}}
=
\mbox{diag}(\widetilde{m}^2_{L_1},\widetilde{m}^2_{L_2},\widetilde{m}^2_{L_3},\widetilde{m}^2_{L_4}),
\end{align}
with $m_{L_i} (i=1,2)$ and $\widetilde{m}^2_{L_i}(i=1,2,3,4)$  being real and non-negative. The mass matrices for messenger quark/squark can be diagonalized in the same way.

Now we are ready to calculate $\mu/B_\mu$-term and soft SUSY breaking parameters. After integrating out messenger fields, ${\mathcal{S}}$ and ${\overline{\mathcal{S}}}$, we find
\begin{align}
\mu
&=
\frac{k_uk_d}{(4\pi)^2}\Lambda_1,\label{eq:mu}\\
A_u
&=
\frac{k^2_u}{(4\pi)^2}\Lambda_2,\\
A_d
&=
\frac{k^2_d}{(4\pi)^2}\Lambda_3,\\
B_\mu
&=
\frac{k_uk_d}{(4\pi)^2}\Lambda^2_4,\label{eq:Bmu}\\
\delta m^2_{H_u}
&=
\frac{k^2_u}{(4\pi)^2}\Lambda^2_5,\\
\delta m^2_{H_d}
&=
\frac{k^2_d}{(4\pi)^2}\Lambda^2_6,
\end{align}
where
\begin{align}
\Lambda_1
&
=m_S\sum_{i=1}^4\widetilde{V}_{1i}\widetilde{V}_{3i}\widetilde{F}_i
+\sum_{i=1}^2 m_{L_i}V_{1i}U_{1i}F^{(-)}_i,\\
\Lambda_2
&=
m_L\sum_{i=1}^4\widetilde{V}_{3i}\widetilde{V}_{1i}\widetilde{F}^{(+)}_i
+m_S\sum_{i=1}^4\widetilde{V}_{1i}\widetilde{V}_{1i}\widetilde{F}^{(-)}_i
+M_{1L}\sum_{i=1}^4\widetilde{V}_{4i}\widetilde{V}_{1i}\widetilde{F}^{(+)}_i,\\
\Lambda_3
&=
m_L\sum_{i=1}^4\widetilde{V}_{1i}\widetilde{V}_{3i}\widetilde{F}^{(+)}_i
+
m_S\sum_{i=1}^4\widetilde{V}_{3i}\widetilde{V}_{3i}\widetilde{F}^{(-)}_i
+
M_{2L}\sum_{i=1}^4\widetilde{V}_{2i}\widetilde{V}_{3i}\widetilde{F}^{(+)}_i
,\\
\Lambda^2_4
&
=
M_{1L}M_{2L}\sum_{i=1}^4\widetilde{V}_{4i}\widetilde{V}_{2i}\widetilde{F}^{(-)}_i
+m_LM_{1L}\sum_{i=1}^4\widetilde{V}_{4i}\widetilde{V}_{1i}\widetilde{F}^{(-)}_i
+m_LM_{2L}\sum_{i=1}^4\widetilde{V}_{3i}\widetilde{V}_{2i}\widetilde{F}^{(-)}_i\nonumber\\
&
+m_SM_{1L}\sum_{i=1}^4\widetilde{V}_{4i}\widetilde{V}_{3i}\widetilde{F}^{(+)}_i
+m_SM_{2L}\sum_{i=1}^4\widetilde{V}_{2i}\widetilde{V}_{1i}\widetilde{F}^{(+)}_i
+m_Sm_L\sum_{i=1}^4\widetilde{V}_{3i}\widetilde{V}_{3i}\widetilde{F}^{(+)}_i\nonumber\\
&
+m_Sm_L\sum_{i=1}^4\widetilde{V}_{1i}\widetilde{V}_{1i}\widetilde{F}^{(+)}_i
+(m^2_L+m^2_S)\sum_{i=1}^4\widetilde{V}_{3i}\widetilde{V}_{1i}\widetilde{F}^{(-)}_i
-2m_S\sum_{i=1}^2m_{L_i}V_{1i}U_{1i}F_i,\\
\Lambda^2_5
&=
A(m^2_S)-\frac{1}{2}A(\widetilde{m}^2_{S_1})-\frac{1}{2}A(\widetilde{m}^2_{S_2})
-\sum_{i=1}^4\widetilde{V}_{1i}\widetilde{V}_{1i}A(\widetilde{m}^2_{L_i})\nonumber\\
&
+M^2_{1L}\sum_{i=1}^4\widetilde{V}_{4i}\widetilde{V}_{4i}\widetilde{F}^{(+)}_i
+m^2_L\sum_{i=1}^4\widetilde{V}_{3i}\widetilde{V}_{3i}\widetilde{F}^{(+)}_i
+m^2_S\sum_{i=1}^4\widetilde{V}_{1i}\widetilde{V}_{1i}\widetilde{F}^{(+)}_i\nonumber\\
&
+2m_LM_{1L}\sum_{i=1}^4\widetilde{V}_{3i}\widetilde{V}_{4i}\widetilde{F}^{(+)}_i
+2m_SM_{1L}\sum_{i=1}^4\widetilde{V}_{1i}\widetilde{V}_{4i}\widetilde{F}^{(-)}_i
+2m_Sm_L\sum_{i=1}^4\widetilde{V}_{1i}\widetilde{V}_{3i}\widetilde{F}^{(-)}_i\nonumber\\
&
+\sum_{i=1}^2V_{1i}V_{1i}A(m^2_{L_i})-\sum_{i=1}^2(m^2_{L_i}+m^2_S)V_{1i}V_{1i}F_i,\\
\Lambda^2_6
&=
A(m^2_S)
-\frac{1}{2}A(\widetilde{m}^2_{S_1})
-\frac{1}{2}A(\widetilde{m}^2_{S_2})
-\sum_{i=1}^4\widetilde{V}_{3i}\widetilde{V}_{3i}A(\widetilde{m}^2_{L_i})\nonumber\\
&
+M^2_{2L}\sum_{i=1}^4\widetilde{V}_{2i}\widetilde{V}_{2i}\widetilde{F}^{(S)}_i
+m^2_L\sum_{i=1}^4\widetilde{V}_{1i}\widetilde{V}_{1i}\widetilde{F}^{(+)}_i
+m^2_S\sum_{i=1}^4\widetilde{V}_{3i}\widetilde{V}_{3i}\widetilde{F}^{(+)}_i\nonumber\\
&
+2m_LM_{2L}\sum_{i=1}^4\widetilde{V}_{1i}\widetilde{V}_{2i}\widetilde{F}^{(+)}_i
+2m_SM_{2L}\sum_{i=1}^4\widetilde{V}_{3i}\widetilde{V}_{2i}\widetilde{F}^{(-)}_i
+2m_Sm_L\sum_{i=1}^4\widetilde{V}_{3i}\widetilde{V}_{1i}\widetilde{F}^{(-)}_i\nonumber\\
&
+\sum_{i=1}^2U_{1i}U_{1i}A(m^2_{L_i})
-\sum_{i=1}^2(m^2_{L_i}+m^2_S)U_{1i}U_{1i}F_i.
\end{align}
Here $F$, $\widetilde{F}$ and $A$ denote the finite one-loop functions which are defined as
\begin{align}
&A(m^2)
=
-m^2\ln m^2,\\
&F_i
=
F_0(m_S, m_{L_i}),\\
&\widetilde{F}_i
=
F_0(m_S, \widetilde{m}_{L_i}),\\
&F^{(\pm)}_i
=
\frac{1}{2}\left[F_0(\widetilde{m}_{S_1}, m_{L_i})\pm F_0(\widetilde{m}_{S_2}, m_{L_i})\right],\\
&\widetilde{F}^{(\pm)}_i
=
\frac{1}{2}\left[F_0(\widetilde{m}_{S_1}, \widetilde{m}_{L_i})\pm F_0(\widetilde{m}_{S_2}, \widetilde{m}_{L_i})\right],
\end{align}
with $m_S=c_S\phi_Z$, $\widetilde{m}^2_{S_1}=m^2_S-c_L\mu^2_Z$, $\widetilde{m}^2_{S_2}=m^2_S+c_L\mu^2_Z$ and
\begin{align}
F_0(m_1,m_2)
&=
\frac{m^2_1}{m^2_1-m^2_2}\ln m^2_1-\frac{m^2_2}{m^2_1-m^2_2}\ln m^2_2.
\end{align}

\end{document}